\documentclass[final,3p,times,twocolumn]{elsarticle}

\usepackage{amssymb}
\usepackage{graphicx,psfrag,color}
\usepackage{dcolumn}
\usepackage{bm}
\usepackage{mathbbol}
\usepackage[T1]{fontenc}

\usepackage{cuted}

\usepackage{amsthm}

\usepackage{amsmath}

\journal{Physics Letters A}

\begin{document}

\begin{frontmatter}



\title{Spin-1 teleportation-based quantum state tomography}




\author{Gustavo Rigolin}
\affiliation{organization={Departamento de F\'isica, Universidade Federal de
S\~ao Carlos, 13565-905, S\~ao Carlos, SP, Brazil}
}

\date{\today}

\begin{abstract}
We show that the teleportation-based quantum state tomography (QST) protocol, 
originally built to reconstruct qubits (spin-1/2 systems), can be extended to deal 
with qutrits (spin-1 systems) as well. Similarly to the original proposal, only
two resources are needed to implement the spin-1 teleportation-based QST protocol: (1) 
Alice should be able to implement the analog of Bell measurements for spin-1 systems;
and (2) she should be able to prepare a few different single qutrit states that will
be teleported to Bob.   
\end{abstract}






\end{frontmatter}



\section{Introduction}

The quantum teleportation protocol \cite{ben93,vai94,bra98,bow97,bos98,fur98,pan06}
aims at transferring the wave function (quantum state) describing one system in a
given place to another system at another place. In quantum information theory parlance,
and when dealing with spin-1/2 systems, one usually says that the goal of the 
quantum teleportation protocol is to ``teleport a qubit from Alice to Bob''. 
The main resource needed to its implementation is a \textit{known} maximally entangled state shared between Alice and Bob. It is through this quantum communication channel that the wave function goes from one location to another.  On the other hand, the state to be teleported does not need to be known. The protocol works even if the state to be teleported is completely unknown. 

The teleportation-based quantum state tomography (QST) protocol \cite{rig26}, 
loosely speaking, turns the standard teleportation protocol on its head. Instead of sending unknown states through known ones, Alice teleports known states through an 
\textit{unknown} quantum state to fully reconstruct the latter. 
This reversal transforms a tool for moving unknown quantum information (quantum states) around into a method for exposing it. This idea of 
reversing the roles of what is known and unknown in the quantum teleportation protocol,
simple in hindsight yet profoundly non-obvious, is the fundamental shift in perspective leading to the teleportation-based QST protocol \cite{rig26}.

In this manuscript we extend the teleportation-based QST protocol of Ref. \cite{rig26}, originally designed to deal with qubits (spin-1/2 systems), to spin-1 systems (qutrits).  We show that we can fully determine an arbitrary two-qutrit state by modifying the teleportation protocol for spin-1 systems given in Refs. \cite{zei19,guo20}. 
We also show how one can generalize the spin-1 teleportation-based QST protocol 
to reconstruct arbitrary $n$-qutrit density matrices. As a bonus, we also provide 
a protocol to reconstruct a single qutrit by means of Bell measurements alone.
Before moving on, we should mention that Refs. \cite{vog89,leo95,whi99,jam01,ari01,bri04,moh06,moh07,moh08,cra10,gro10,tot10,chr12,bau13,bau13b,lan17,tor18} describe important and ground breaking QST protocols that do not
use the teleportation protocol as a key ingredient.

\section{Reconstructing two-qutrits}

The quantum state we want to completely determine is in general a mixed state, with a 
pure state being a special case of the latter. Thus, we must theoretically work with the teleportation protocol expressed in the language of density
matrices \cite{rig26,rig15,pav23}, which are the proper mathematical objects we need to deal with either pure or mixed states \cite{nie00}.

\begin{figure}[!ht]
\centering
\includegraphics[width=5cm]{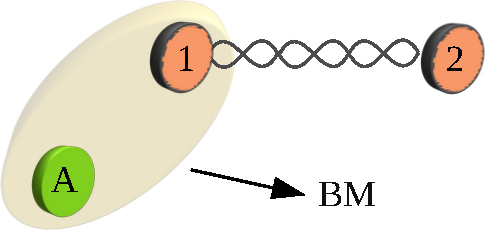}
\caption{\label{fig1}(color online) 
Schematic view of the teleportation-based QST protocol, whose goal is to fully
reconstruct the two-qutrit state $\rho_{12}$ (orange disks $1$ and $2$).
One run of the protocol works as follows.
First, Alice prepares an input state $\rho_A$ (green disk $A$), choosing among nine 
possible input states as explained in the main text. Second,   
she implements onto qutrits $A$ and $1$ a Bell measurement (BM), informing Bob of 
her measurement result and which qutrit $A$ she teleported. After receiving these two pieces of information, Bob is sure that $\varrho_2$ [Eq.~(\ref{stepE})] describes his qutrit. This procedure is repeated several times using the nine different input states 
that Alice should be able to prepare. In the end, Bob can fully reconstruct
the initial state describing qutrits $1$ and $2$, i.e. $\rho_{12}$,  with the knowledge of all states $\varrho_2$ associated with each different input state $\rho_A$ teleported by Alice.}
\end{figure}

As such, we represent the quantum state shared between our two parties, Alice and Bob, by the density matrix $\rho_{12}$. Qutrit $1$ is with Alice and qutrit $2$ with Bob. The state $\rho_{12}$ is unknown and our goal is to fully determine it by teleporting 
known states from Alice to Bob. The known state that Alice teleports to Bob is 
a pure state $|\psi\rangle=\alpha|0\rangle + \beta|1\rangle+  \gamma|2\rangle$,
whose density matrix is $\rho_{A}=|\psi\rangle\langle \psi|$. 
We also assume that the state $|\psi\rangle$ is normalized, i.e., 
$|\alpha|^2+|\beta|^2+|\gamma|^2=1$. See Fig.~\ref{fig1} for a step by step description of the teleportation protocol.

The state describing all qutrits before the start of any run of the teleportation protocol is 
\begin{equation}
\rho = \rho_{A} \otimes \rho_{12} = 
\left(
\begin{array}{ccc}
|\alpha|^2 & \alpha\beta^* & \alpha\gamma^* \\
\alpha^*\beta & |\beta|^2 & \beta\gamma^* \\
\alpha^*\gamma & \beta^*\gamma & |\gamma|^2
\end{array}
\right) \otimes \rho_{12},
\label{stepA}
\end{equation}
where $*$ denotes complex conjugation, and after a single run
of the protocol, Bob's qutrit is given by \cite{rig15,pav23}
\begin{equation}
\varrho_{2}=\frac{1}{Q_j}[U_jTr_{A,1}(P_j \rho P_j)U_j^\dagger].
\label{stepD}
\end{equation}
In Eq.~(\ref{stepD}) the notation $Tr_{A,1}$ means the partial trace over Alice's qutrits, $j$ denotes the Bell measurement (BM) outcome
obtained by Alice ($j=1, \ldots, 9$), and $P_j$ 
gives those projectors describing the nine possible BM results,
\begin{equation}
P_{j} = |B_j\rangle \langle B_j|,
\label{projectorB}  
\end{equation}
with the spin-1 Bell states being \cite{zei19,guo20},
\begin{eqnarray}
|B_1\rangle &=& (|00\rangle + |11\rangle + |22\rangle)/\sqrt{3}, \\
|B_2\rangle &=& (|00\rangle + e^{2\pi i/3}|11\rangle + e^{4\pi i/3}|22\rangle)/\sqrt{3}, \\
|B_3\rangle &=& (|00\rangle + e^{4\pi i/3}|11\rangle + e^{2\pi i/3}|22\rangle)/\sqrt{3}, \\
|B_4\rangle &=& (|01\rangle + |12\rangle + |20\rangle)/\sqrt{3}, \\
|B_5\rangle &=& (|01\rangle + e^{2\pi i/3}|12\rangle + e^{4\pi i/3}|20\rangle)/\sqrt{3}, \\
|B_6\rangle &=& (|01\rangle + e^{4\pi i/3}|12\rangle + e^{2\pi i/3}|20\rangle)/\sqrt{3}, \\
|B_7\rangle &=& (|02\rangle + |10\rangle + |21\rangle)/\sqrt{3}, \\
|B_8\rangle &=&(|02\rangle + e^{2\pi i/3}|10\rangle + e^{4\pi i/3}|21\rangle)/\sqrt{3}, \\
|B_9\rangle &=& (|02\rangle + e^{4\pi i/3}|10\rangle + e^{2\pi i/3}|21\rangle)/\sqrt{3}.\label{BellB}
\end{eqnarray}

Alice's chance of measuring a Bell state $|B_j\rangle$ depends on
the input state and on the shared state between Alice and Bob. This probability
is given by \cite{rig15,pav23}
\begin{equation}
Q_j=Q_j(\alpha,\beta,\gamma) = Tr({P_j \rho}),
\label{prob}
\end{equation}
where we highlight its dependence on the input state.

The unitary operation (correction) Bob should implement onto his qutrit is $U_j$. It depends on Alice's BM result $|B_j\rangle$. In order to apply the appropriate unitary operation, Bob must receive from Alice the information concerning the result of her BM. These nine unitary corrections are given by $3\times 3$ unitary matrices and can be seen in Refs. \cite{zei19,guo20}. For the teleportation-based QST protocol, 
they are not needed since we can work directly with Bob's ``raw'' (uncorrected) state \cite{rig26}. From now on, thus, we simply set those corrections to $\mathbb{1}$, the 
$3\times 3$ identity matrix. With that in mind, at the end of each run of the  teleportation-based QST protocol, Bob's state is given by 
\begin{equation}
\varrho_{2} =  \frac{Tr_{A,1}(P_j \rho P_j)}{Q_j}
=
\left(
\begin{array}{ccc}
b_{11} & b_{12} & b_{13} \\
b_{12}^* & b_{22} & b_{23} \\
b_{13}^* & b_{23}^* & b_{33}
\end{array}
\right). 
\label{stepE}
\end{equation}
Equation (14) plays a crucial role in the implementation of the teleportation-based QST protocol. We assume that Bob is able to experimentally reconstruct his single-qutrit
density matrix. In other words, at the end of the teleportation protocol, Bob can determine all the elements $b_{ij}$ of his qutrit state using any standard QST method \cite{vog89,leo95,whi99,jam01,ari01,bri04,moh06,moh07,moh08,cra10,gro10,tot10,chr12,bau13,bau13b,lan17,tor18} or the Bell-state-measurement-only QST protocol \cite{rig26},
which we extend here to the qutrit case. The coefficients $b_{ij}$ depend on the matrix
elements of the unknown shared state $\rho_{12}$ as well as on the input state 
teleported by Alice. As we show below, by teleporting only nine different pure input
states, we obtain a consistent system of linear equations for the matrix elements of 
$\rho_{12}$. Solving this system yields these matrix elements as functions of $b_{ij}$, i.e., the matrix elements of Bob’s state. In what follows, it will be convenient to 
work with the unnormalized matrix elements of $\varrho_2$, 
\begin{equation}
\tilde{b}_{ij}=\tilde{b}_{ij}(\alpha,\beta,\gamma) =  Q_j(\alpha,\beta,\gamma)
b_{ij}(\alpha,\beta,\gamma), \label{btilde}
\end{equation}
where we highlight for later convenience the dependence of $\tilde{b}_{ij}, b_{ij}$, and $Q_j$ on the input state.

Noting that a physical density matrix should be positive definite and Hermitian, the most general two-qutrit state is
\begin{eqnarray}
\hspace{-1cm}\rho_{12} \!\!\!\! &=&\!\!\!\! \left(\!\!
\begin{array}{ccccccccc}
m_{11} \hspace{-.1cm}&\hspace{-.1cm} m_{12} \hspace{-.1cm}&\hspace{-.1cm} m_{13} \hspace{-.1cm}&\hspace{-.1cm} m_{14} \hspace{-.1cm}&\hspace{-.1cm} m_{15} \hspace{-.1cm}&\hspace{-.1cm} m_{16} \hspace{-.1cm}&\hspace{-.1cm} m_{17} \hspace{-.1cm}&\hspace{-.1cm} m_{18} \hspace{-.1cm}&\hspace{-.1cm} m_{19} \\
m_{12}^* \hspace{-.1cm}&\hspace{-.1cm} m_{22} \hspace{-.1cm}&\hspace{-.1cm} m_{23} \hspace{-.1cm}&\hspace{-.1cm} m_{24} \hspace{-.1cm}&\hspace{-.1cm} m_{25} \hspace{-.1cm}&\hspace{-.1cm} m_{26} \hspace{-.1cm}&\hspace{-.1cm} m_{27} \hspace{-.1cm}&\hspace{-.1cm} m_{28} \hspace{-.1cm}&\hspace{-.1cm} m_{29} \\
m_{13}^* \hspace{-.1cm}&\hspace{-.1cm} m_{23}^* \hspace{-.1cm}&\hspace{-.1cm} m_{33} \hspace{-.1cm}&\hspace{-.1cm} m_{34} \hspace{-.1cm}&\hspace{-.1cm} m_{35} \hspace{-.1cm}&\hspace{-.1cm} m_{36} \hspace{-.1cm}&\hspace{-.1cm} m_{37} \hspace{-.1cm}&\hspace{-.1cm} m_{38} \hspace{-.1cm}&\hspace{-.1cm} m_{39} \\
m_{14}^* \hspace{-.1cm}&\hspace{-.1cm} m_{24}^* \hspace{-.1cm}&\hspace{-.1cm} m_{34}^* \hspace{-.1cm}&\hspace{-.1cm} m_{44} \hspace{-.1cm}&\hspace{-.1cm} m_{45} \hspace{-.1cm}&\hspace{-.1cm} m_{46} \hspace{-.1cm}&\hspace{-.1cm} m_{47} \hspace{-.1cm}&\hspace{-.1cm} m_{48} \hspace{-.1cm}&\hspace{-.1cm} m_{49} \\
m_{15}^* \hspace{-.1cm}&\hspace{-.1cm} m_{25}^* \hspace{-.1cm}&\hspace{-.1cm} m_{35}^* \hspace{-.1cm}&\hspace{-.1cm} m_{45}^* \hspace{-.1cm}&\hspace{-.1cm} m_{55} \hspace{-.1cm}&\hspace{-.1cm} m_{56} \hspace{-.1cm}&\hspace{-.1cm} m_{57} \hspace{-.1cm}&\hspace{-.1cm} m_{58} \hspace{-.1cm}&\hspace{-.1cm} m_{59} \\
m_{16}^* \hspace{-.1cm}&\hspace{-.1cm} m_{26}^* \hspace{-.1cm}&\hspace{-.1cm} m_{36}^* \hspace{-.1cm}&\hspace{-.1cm} m_{46}^* \hspace{-.1cm}&\hspace{-.1cm} m_{56}^* \hspace{-.1cm}&\hspace{-.1cm} m_{66} \hspace{-.1cm}&\hspace{-.1cm} m_{67} \hspace{-.1cm}&\hspace{-.1cm} m_{68} \hspace{-.1cm}&\hspace{-.1cm} m_{69} \\
m_{17}^* \hspace{-.1cm}&\hspace{-.1cm} m_{27}^* \hspace{-.1cm}&\hspace{-.1cm} m_{37}^* \hspace{-.1cm}&\hspace{-.1cm} m_{47}^* \hspace{-.1cm}&\hspace{-.1cm} m_{57}^* \hspace{-.1cm}&\hspace{-.1cm} m_{67}^* \hspace{-.1cm}&\hspace{-.1cm} m_{77} \hspace{-.1cm}&\hspace{-.1cm} m_{78} \hspace{-.1cm}&\hspace{-.1cm} m_{79} \\
m_{18}^* \hspace{-.1cm}&\hspace{-.1cm} m_{28}^* \hspace{-.1cm}&\hspace{-.1cm} m_{38}^* \hspace{-.1cm}&\hspace{-.1cm} m_{48}^* \hspace{-.1cm}&\hspace{-.1cm} m_{58}^* \hspace{-.1cm}&\hspace{-.1cm} m_{68}^* \hspace{-.1cm}&\hspace{-.1cm} m_{78}^* \hspace{-.1cm}&\hspace{-.1cm} m_{88} \hspace{-.1cm}&\hspace{-.1cm} m_{89} \\
m_{19}^* \hspace{-.1cm}&\hspace{-.1cm} m_{29}^* \hspace{-.1cm}&\hspace{-.1cm} m_{39}^* \hspace{-.1cm}&\hspace{-.1cm} m_{49}^* \hspace{-.1cm}&\hspace{-.1cm} m_{59}^* \hspace{-.1cm}&\hspace{-.1cm} m_{69}^* \hspace{-.1cm}&\hspace{-.1cm} m_{79}^* \hspace{-.1cm}&\hspace{-.1cm} m_{89}^* \hspace{-.1cm}&\hspace{-.1cm} m_{99}
\end{array}
\!\!\right)\!\! \nonumber \\
\!\hspace{-.1cm}&\hspace{-.1cm}=&\!\!\!\! \left(\!\!
\begin{array}{ccc}
A_{3\times3} & B_{3\times3} & C_{3\times3} \\
B^*_{3\times3} & E_{3\times3} & F_{3\times3} \\
C^*_{3\times3} & F^*_{3\times3} & I_{3\times3}
\end{array}
\!\!\right).\label{rho12}
\end{eqnarray}
This is the quantum state we want to reconstruct using the teleportation-based 
QST protocol. In Eq.~(\ref{rho12}) $m_{jj}$ are real positive numbers obeying
the normalization condition, namely, $\sum_{j}m_{jj}=1$.  We have $81$ real parameters to determine in order to reconstruct $\rho_{12}$ ($80$ if the normalization condition
is used). Indeed, we have $9$ real diagonal elements to determine and $72$ real numbers to determine associated with the real and imaginary parts of the off-diagonal terms.
For ease of analysis, we have divided the $9\times 9$ matrix into nine blocks of 
$3\times 3$ matrices, as shown in the last line of Eq.~(\ref{rho12}). Note that 
only six blocks are independent.

Employing Eqs.~(\ref{stepA}), (\ref{projectorB}), and (\ref{rho12}),
a straightforward but long calculation using Eq.~(\ref{prob}) show that Alice's probability to measure 
the Bell state $|B_j\rangle$ is
%
%
\begin{eqnarray}
\hspace{-.5cm}Q_1(\alpha,\beta,\gamma) &=&\frac{1}{3} \left\{ \left( m_{11} + m_{22} + m_{33} \right) |\alpha|^2 \right. \nonumber \\
&&+ \left( m_{44} + m_{55} + m_{66} \right) |\beta|^2 \nonumber \\
&&+\left( m_{77} + m_{88} + m_{99} \right)|\gamma|^2 \nonumber \\  
&& + 2 \Re \left[ \alpha \beta^*\left( m_{14} + m_{25} + m_{36} \right)  \right] 
\nonumber \\
&&+ 2 \Re \left[ \alpha \gamma^*\left( m_{17} + m_{28} + m_{39} \right)  \right]
\nonumber \\ 
&&\left.+ 2 \Re \left[ \beta \gamma^*\left( m_{47} + m_{58} + m_{69} \right)  \right] \right\}, \label{q1}\\
\hspace{-.5cm}Q_2(\alpha, \beta, \gamma) &=& Q_1 \left( \alpha, e^{i \frac{4\pi}{3}} \beta, e^{i \frac{2\pi}{3}} \gamma \right), \label{q2} \\
\hspace{-.5cm}Q_3(\alpha, \beta, \gamma) &=& Q_1 \left( \alpha, e^{i \frac{2\pi}{3}} \beta, e^{i \frac{4\pi}{3}} \gamma \right),\\
\hspace{-.5cm}Q_4(\alpha, \beta, \gamma) &=& Q_1 \left( \gamma, \alpha, \beta \right),\\
\hspace{-.5cm}Q_5(\alpha, \beta, \gamma) &=& Q_1 \left( \gamma, e^{i \frac{4\pi}{3}} \alpha, e^{i \frac{2\pi}{3}} \beta \right),\\
\hspace{-.5cm}Q_6(\alpha, \beta, \gamma) &=& Q_1 \left( \gamma, e^{i \frac{2\pi}{3}} \alpha, e^{i \frac{4\pi}{3}} \beta \right),\\
\hspace{-.5cm}Q_7(\alpha, \beta, \gamma) &=& Q_1 \left( \beta, \gamma, \alpha \right),\\
\hspace{-.5cm}Q_8(\alpha, \beta, \gamma) &=& Q_1 \left( \beta, e^{i \frac{4\pi}{3}} \gamma, e^{i \frac{2\pi}{3}} \alpha \right),\\
\hspace{-.5cm}Q_9(\alpha, \beta, \gamma) &=& Q_1 \left( \beta, e^{i \frac{2\pi}{3}} \gamma, e^{i \frac{4\pi}{3}} \alpha \right), \label{q9}
\end{eqnarray}
where $\Re[z]$ is the real part of the complex number $z$.

Analogously, using Eqs.~(\ref{stepE}) and (\ref{btilde}) we obtain Bob's state at the end of the teleportation protocol. For a given input state, there are nine possible states with Bob, depending on which Bell state $|B_j\rangle$ Alice measures. And each state has nine matrix elements,
with only six independent due to the Hermiticity condition, i.e., 
$\tilde{b}_{ij}=\tilde{b}^*_{ji}$. Assuming Alice measures the state $|B_1\rangle$, those six matrix coefficients characterizing Bob's state are
\begin{align}
\hspace{-.2cm}\tilde{b}_{11}(\alpha,\beta,\gamma) = 
\frac{1}{3}\left[m_{11}|\alpha|^2 + m_{44}|\beta|^2 + m_{77}|\gamma|^2 \right.
\nonumber \\
\hspace{-.2cm}+\left. 2\Re\!\left(\alpha \beta^* m_{14}+\alpha \gamma^* m_{17}+\beta \gamma^* m_{47}\right)
\right], \label{eqb11} \\
\hspace{-.2cm}\tilde{b}_{12}(\alpha,\beta,\gamma) = 
\frac{1}{3}\left[m_{12}|\alpha|^2 + m_{45}|\beta|^2 + m_{78}|\gamma|^2
+ \alpha \beta^* m_{15}  
\right. \nonumber \\
\hspace{-.2cm}+\left.\alpha \gamma^* m_{18} + \beta \gamma^* m_{48} + \beta \alpha^* m_{24}^* + \gamma \alpha^* m_{27}^* + \gamma \beta^* m_{57}^*
\right], \label{eqb12}\\
\hspace{-.2cm}\tilde{b}_{13}(\alpha,\beta,\gamma) = 
\frac{1}{3}\left[ m_{13}|\alpha|^2 + m_{46}|\beta|^2 + m_{79}|\gamma|^2
+ \alpha \beta^* m_{16} 
\right. \nonumber \\
\hspace{-.2cm}+\left. \alpha \gamma^* m_{19} + \beta \gamma^* m_{49} + \beta \alpha^* m_{34}^* + \gamma \alpha^* m_{37}^* + \gamma \beta^* m_{67}^*
\right], \\
\hspace{-.2cm}\tilde{b}_{22}(\alpha,\beta,\gamma)=\frac{1}{3}\left[
m_{22}|\alpha|^2 + m_{55}|\beta|^2 + m_{88}|\gamma|^2 \right. \nonumber \\
\hspace{-.2cm}+\left.  2\Re\!\left(\alpha \beta^* m_{25}+\alpha \gamma^* m_{28}+\beta \gamma^* m_{58}\right)
\right], \label{eqb22} \\
\hspace{-.2cm}\tilde{b}_{23}(\alpha,\beta,\gamma) = \frac{1}{3}\left[
m_{23}|\alpha|^2 + m_{56}|\beta|^2 + m_{89}|\gamma|^2
+ \alpha \beta^* m_{26}  
 \right. \nonumber \\
\hspace{-.2cm}+ \left. \alpha \gamma^* m_{29} + \beta \gamma^* m_{59} + \beta \alpha^* m_{35}^* + \gamma \alpha^* m_{38}^* + \gamma \beta^* m_{68}^*
\right],\\
\hspace{-.2cm}\tilde{b}_{33}(\alpha,\beta,\gamma) =\frac{1}{3}\left[
m_{33}|\alpha|^2 + m_{66}|\beta|^2 + m_{99}|\gamma|^2 \right. \nonumber \\
\hspace{-.2cm}+\left. 2\Re\!\left(\alpha \beta^* m_{36}+\alpha \gamma^* m_{39}+\beta \gamma^* m_{69}\right)
\right]. \label{eqb33}
\end{align}

For the other eight possible BM outcomes that Alice might obtain, a direct calculation
shows that Bob's states are related to each other in exactly the same way as the probabilities of obtaining each Bell state. In other words,
\begin{eqnarray}
\varrho_{2,B_2}(\alpha, \beta, \gamma) &=& \varrho_{2,B_1} \left( \alpha, e^{i \frac{4\pi}{3}} \beta, e^{i \frac{2\pi}{3}} \gamma \right), \label{rb2}\\
\varrho_{2,B_3}(\alpha, \beta, \gamma) &=& \varrho_{2,B_1} \left( \alpha, e^{i \frac{2\pi}{3}} \beta, e^{i \frac{4\pi}{3}} \gamma \right),\\
\varrho_{2,B_4}(\alpha, \beta, \gamma) &=& \varrho_{2,B_1} \left( \gamma, \alpha, \beta \right),\\
\varrho_{2,B_5}(\alpha, \beta, \gamma) &=& \varrho_{2,B_1} \left( \gamma, e^{i \frac{4\pi}{3}} \alpha, e^{i \frac{2\pi}{3}} \beta \right), \\
\varrho_{2,B_6}(\alpha, \beta, \gamma) &=& \varrho_{2,B_1} \left( \gamma, e^{i \frac{2\pi}{3}} \alpha, e^{i \frac{4\pi}{3}} \beta \right),\\
\varrho_{2,B_7}(\alpha, \beta, \gamma) &=& \varrho_{2,B_1} \left( \beta, \gamma, \alpha \right), \\
\varrho_{2,B_8}(\alpha, \beta, \gamma) &=& \varrho_{2,B_1} \left( \beta, e^{i \frac{4\pi}{3}} \gamma, e^{i \frac{2\pi}{3}} \alpha \right),\\
\varrho_{2,B_9}(\alpha, \beta, \gamma) &=& \varrho_{2,B_1} \left( \beta, e^{i \frac{2\pi}{3}} \gamma, e^{i \frac{4\pi}{3}} \alpha \right), \label{rb9}
\end{eqnarray}
where $\varrho_{2,B_j}$ means the state with Bob assuming Alice measures the Bell
state $|B_j\rangle$.

We now assume, for definiteness and without losing in generality, that Alice has
measured the Bell state $|B_1\rangle$. Of course, the same analysis is valid if she 
measures any Bell state. Or, equivalently, if she measures a different Bell state
than $|B_1\rangle$, Bob can post-process his data by applying the rules given 
by Eqs.~(\ref{q2})-(\ref{q9}) and (\ref{rb2})-(\ref{rb9}). In this way, he will
effectively be dealing with $\varrho_{2,B_1}$ and can use the expressions below to obtain the matrix elements of $\rho_{12}$.

If Alice teleports the state $|\psi\rangle=|0\rangle$ 
($\alpha,\beta,\gamma=1,0,0$), a direct inspection at Eqs.~(\ref{eqb11})-(\ref{eqb33})
tells us that we can determine 
the coefficients $m_{11},m_{22},m_{33}, m_{12}, m_{13}, m_{23}$. Similarly, if she
teleports $|\psi\rangle=|1\rangle$ ($\alpha,\beta,\gamma=0,1,0$) we
get $m_{44},m_{55}, m_{66}, m_{45}, m_{46}, m_{56}$ and if she teleports 
the state $|\psi\rangle=|2\rangle$ ($\alpha,\beta,\gamma=0,0,1$) we can determine 
$m_{77},m_{88},m_{99}, m_{78}, m_{79}, m_{89}$. Note that by teleporting the state 
$|0\rangle$ we can fully reconstruct the block matrix $A_{3\times3}$, by teleporting 
$|1\rangle$ we obtain $E_{3\times3}$, and by teleporting $|2\rangle$ 
we get $I_{3\times3}$.
In other words, by teleporting each one of the states that span
the standard basis $\{|0\rangle,|1\rangle,|2\rangle\}$, we obtain the block diagonal matrices associated to $\rho_{12}$. So far we have obtained $27$ real parameters
out of the $81$ needed to fully reconstruct $\rho_{12}$. The explicit solution to
this simple system of equations is
\begin{eqnarray}
\mbox{Block} \;\; A_{3\times3}: & &  \nonumber \\
m_{11} &=& 3 \tilde{b}_{11}(1,0,0),\label{1of81} \\
m_{22} &=& 3 \tilde{b}_{22}(1,0,0), \\
m_{33} &=& 3 \tilde{b}_{33}(1,0,0), \\
m_{12} &=& 3 \tilde{b}_{12}(1,0,0), \\
m_{13} &=& 3 \tilde{b}_{13}(1,0,0), \\
m_{23} &=& 3 \tilde{b}_{23}(1,0,0), \\
\mbox{Block} \;\; E_{3\times3}: & &  \nonumber \\
m_{44} &=& 3 \tilde{b}_{44}(0,1,0), \\
m_{55} &=& 3 \tilde{b}_{55}(0,1,0), \\
m_{66} &=& 3 \tilde{b}_{66}(0,1,0), \\
m_{45} &=& 3 \tilde{b}_{12}(0,1,0), \\
m_{46} &=& 3 \tilde{b}_{13}(0,1,0), \\
m_{56} &=& 3 \tilde{b}_{23}(0,1,0), 
\end{eqnarray}
\begin{eqnarray}
\mbox{Block} \;\; I_{3\times3}: & &  \nonumber \\
m_{77} &=& 3 \tilde{b}_{77}(0,0,1), \\
m_{88} &=& 3 \tilde{b}_{88}(0,0,1), \\
m_{99} &=& 3 \tilde{b}_{99}(0,0,1), \\
m_{78} &=& 3 \tilde{b}_{12}(0,0,1), \\
m_{79} &=& 3 \tilde{b}_{13}(0,0,1), 
\end{eqnarray}
\begin{eqnarray}
m_{89} &=& 3 \tilde{b}_{23}(0,0,1). \label{27of81}
\end{eqnarray}

On the other hand, by teleporting  
$|\psi\rangle$ $=(|0\rangle+|1\rangle)/\sqrt{2}$  
($\alpha,\beta,\gamma=1/\sqrt{2},$ $1/\sqrt{2},0$) and 
$|\psi\rangle=$ $(|0\rangle+i|1\rangle)/\sqrt{2}$  
($\alpha,\beta,\gamma=1/\sqrt{2},i/\sqrt{2},0$), we can solve 
Eqs.~(\ref{eqb11})-(\ref{eqb33}) for $m_{14},m_{15},m_{16},m_{24},m_{25},m_{26},m_{34},
m_{35},m_{36}$, all coefficients of block $B_{3\times3}$,
\begin{eqnarray}
\hspace{-1.5cm}m_{14} \hspace{-.3cm}&=&\hspace{-.3cm} -\frac{3}{2}(1 + i) [\tilde{b}_{11}(1,0,0) + 
\tilde{b}_{11}(0,1,0) ] \nonumber \\ 
\hspace{-.3cm}&+&\hspace{-.3cm} 3 \tilde{b}_{11}(1/\!\sqrt{2}, 1/\!\sqrt{2}, 0) \label{28of81}
+ 3i \tilde{b}_{11}( 1/\!\sqrt{2},i/\!\sqrt{2},0),\\
\hspace{-1.5cm}m_{15} \hspace{-.3cm}&=&\hspace{-.3cm} -\frac{3}{2}(1 + i) [\tilde{b}_{12}(1,0,0) + 
\tilde{b}_{12}(0,1,0) ] \nonumber \\ 
\hspace{-.3cm}&+&\hspace{-.3cm} 3 \tilde{b}_{12}(1/\!\sqrt{2}, 1/\!\sqrt{2}, 0)
+ 3i \tilde{b}_{12}( 1/\!\sqrt{2},i/\!\sqrt{2},0),\\
\hspace{-1.5cm}m_{16} \hspace{-.3cm}&=&\hspace{-.3cm} -\frac{3}{2}(1 + i) [\tilde{b}_{13}(1,0,0) + 
\tilde{b}_{13}(0,1,0) ] \nonumber \\ 
\hspace{-.3cm}&+&\hspace{-.3cm} 3 \tilde{b}_{13}(1/\!\sqrt{2}, 1/\!\sqrt{2}, 0)
+ 3i \tilde{b}_{13}( 1/\!\sqrt{2},i/\!\sqrt{2},0), \\
\hspace{-1.5cm}m_{24} \hspace{-.3cm}&=&\hspace{-.3cm} -\frac{3}{2}(1 + i) [\tilde{b}_{21}(1,0,0) + 
\tilde{b}_{21}(0,1,0) ] \nonumber \\ 
\hspace{-.3cm}&+&\hspace{-.3cm} 3 \tilde{b}_{21}(1/\!\sqrt{2}, 1/\!\sqrt{2}, 0)
+ 3i \tilde{b}_{21}( 1/\!\sqrt{2},i/\!\sqrt{2},0), \\
\hspace{-1.5cm}m_{25} \hspace{-.3cm}&=&\hspace{-.3cm} -\frac{3}{2}(1 + i)[\tilde{b}_{22}(1,0,0) + \tilde{b}_{22}(0,1,0) ]
\nonumber \\
\hspace{-.3cm}&+&\hspace{-.3cm} 3 \tilde{b}_{22}( 1/\!\sqrt{2}, 1/\!\sqrt{2}, 0 ) 
+ 3i\tilde{b}_{22}(1/\!\sqrt{2}, i/\!\sqrt{2}, 0), \\
\hspace{-1.5cm}m_{26} \hspace{-.3cm}&=&\hspace{-.3cm} -\frac{3}{2}(1 + i) [\tilde{b}_{23}(1,0,0) + 
\tilde{b}_{23}(0,1,0) ] \nonumber \\ 
\hspace{-.3cm}&+&\hspace{-.3cm} 3 \tilde{b}_{23}(1/\!\sqrt{2}, 1/\!\sqrt{2}, 0)
+ 3i \tilde{b}_{23}( 1/\!\sqrt{2},i/\!\sqrt{2},0),\\
\hspace{-1.5cm}m_{34} \hspace{-.3cm}&=&\hspace{-.3cm} -\frac{3}{2}(1 + i) [\tilde{b}_{31}(1,0,0) + 
\tilde{b}_{31}(0,1,0) ] \nonumber \\ 
\hspace{-.3cm}&+&\hspace{-.3cm} 3 \tilde{b}_{31}(1/\!\sqrt{2}, 1/\!\sqrt{2}, 0)
+ 3i \tilde{b}_{31}( 1/\!\sqrt{2},i/\!\sqrt{2},0),\\
\hspace{-1.5cm}m_{35} \hspace{-.3cm}&=&\hspace{-.3cm} -\frac{3}{2}(1 + i) [\tilde{b}_{32}(1,0,0) + 
\tilde{b}_{32}(0,1,0) ] \nonumber \\ 
\hspace{-.3cm}&+&\hspace{-.3cm} 3 \tilde{b}_{32}(1/\!\sqrt{2}, 1/\!\sqrt{2}, 0)
+ 3i \tilde{b}_{32}( 1/\!\sqrt{2},i/\!\sqrt{2},0),\\
\hspace{-1.5cm}m_{36} \hspace{-.3cm}&=&\hspace{-.3cm} -\frac{3}{2}(1 + i)[\tilde{b}_{33}(1,0,0) + \tilde{b}_{33}(0,1,0) ]
\nonumber \\
\hspace{-.3cm}&+&\hspace{-.3cm} 3 \tilde{b}_{33}( 1/\!\sqrt{2}, 1/\!\sqrt{2}, 0 ) 
+ 3i\tilde{b}_{33}(1/\!\sqrt{2}, i/\!\sqrt{2}, 0).
\label{45of81}
\end{eqnarray}
Note that we have also used Eqs.~(\ref{1of81})-(\ref{27of81}) to express 
Eqs.~(\ref{27of81})-(\ref{45of81}) in terms of Bob's measured quantities.

If Alice teleports $|\psi\rangle$ $=(|0\rangle+|2\rangle)/\sqrt{2}$  
($\alpha,\beta,\gamma=1/\sqrt{2}$, $0$, $1/\sqrt{2}$) and 
$|\psi\rangle=$ $(|0\rangle+i|2\rangle)/\sqrt{2}$  
($\alpha,\beta,\gamma=1/\sqrt{2}$, $0$, $i/\sqrt{2}$),  
we can solve Eqs.~(\ref{eqb11})-(\ref{eqb33}) to get $m_{17},m_{18},m_{19},m_{27},m_{28},m_{29},m_{37},
m_{38},m_{39}$, the coefficients of block $C_{3\times3}$,
\begin{eqnarray}
\hspace{-1.2cm}m_{17} \hspace{-.3cm}&=&\hspace{-.3cm} - \frac{3}{2}(1+i)
[\tilde{b}_{11}(1,0,0) + \tilde{b}_{11}(0,0,1) ]
\nonumber \\
\hspace{-.3cm}&+&\hspace{-.3cm} 3\tilde{b}_{11}(1/\!\sqrt{2},0,1/\!\sqrt{2}) +
3i\tilde{b}_{11}(1/\!\sqrt{2},0,i/\!\sqrt{2}), \label{46of81} \\
\hspace{-1.2cm}m_{18} \hspace{-.3cm}&=&\hspace{-.3cm} - \frac{3}{2}(1+i)
[\tilde{b}_{12}(1,0,0) + \tilde{b}_{12}(0,0,1) ]
\nonumber \\
\hspace{-.3cm}&+&\hspace{-.3cm} 3\tilde{b}_{12}(1/\!\sqrt{2},0,1/\!\sqrt{2}) +
3i\tilde{b}_{12}(1/\!\sqrt{2},0,i/\!\sqrt{2}), 
\end{eqnarray}
\begin{eqnarray}
\hspace{-1.2cm}m_{19} \hspace{-.3cm}&=&\hspace{-.3cm} - \frac{3}{2}(1+i)
[\tilde{b}_{13}(1,0,0) + \tilde{b}_{13}(0,0,1) ]
\nonumber \\
\hspace{-.3cm}&+&\hspace{-.3cm} 3\tilde{b}_{13}(1/\!\sqrt{2},0,1/\!\sqrt{2}) +
3i\tilde{b}_{13}(1/\!\sqrt{2},0,i/\!\sqrt{2}), \\
\hspace{-1.2cm}m_{27} \hspace{-.3cm}&=&\hspace{-.3cm} - \frac{3}{2}(1+i)
[\tilde{b}_{21}(1,0,0) + \tilde{b}_{21}(0,0,1) ]
\nonumber \\
\hspace{-.3cm}&+&\hspace{-.3cm} 3\tilde{b}_{21}(1/\!\sqrt{2},0,1/\!\sqrt{2}) +
3i\tilde{b}_{21}(1/\!\sqrt{2},0,i/\!\sqrt{2}), \\
\hspace{-1.2cm}m_{28} \hspace{-.3cm}&=&\hspace{-.3cm} - \frac{3}{2}(1+i)
[\tilde{b}_{22}(1,0,0) + \tilde{b}_{22}(0,0,1) ]
\nonumber \\
\hspace{-.3cm}&+&\hspace{-.3cm} 3\tilde{b}_{22}(1/\!\sqrt{2},0,1/\!\sqrt{2}) +
3i\tilde{b}_{22}(1/\!\sqrt{2},0,i/\!\sqrt{2}), \\
\hspace{-1.2cm}m_{29} \hspace{-.3cm}&=&\hspace{-.3cm} - \frac{3}{2}(1+i)
[\tilde{b}_{23}(1,0,0) + \tilde{b}_{23}(0,0,1) ]
\nonumber \\
\hspace{-.3cm}&+&\hspace{-.3cm} 3\tilde{b}_{23}(1/\!\sqrt{2},0,1/\!\sqrt{2}) +
3i\tilde{b}_{23}(1/\!\sqrt{2},0,i/\!\sqrt{2}), \\
\hspace{-1.2cm}m_{37} \hspace{-.3cm}&=&\hspace{-.3cm} - \frac{3}{2}(1+i)
[\tilde{b}_{31}(1,0,0) + \tilde{b}_{31}(0,0,1) ]
\nonumber \\
\hspace{-.3cm}&+&\hspace{-.3cm} 3\tilde{b}_{31}(1/\!\sqrt{2},0,1/\!\sqrt{2}) +
3i\tilde{b}_{31}(1/\!\sqrt{2},0,i/\!\sqrt{2}), \\
\hspace{-1.2cm}m_{38} \hspace{-.3cm}&=&\hspace{-.3cm} - \frac{3}{2}(1+i)
[\tilde{b}_{32}(1,0,0) + \tilde{b}_{32}(0,0,1) ]
\nonumber \\
\hspace{-.3cm}&+&\hspace{-.3cm} 3\tilde{b}_{32}(1/\!\sqrt{2},0,1/\!\sqrt{2}) +
3i\tilde{b}_{32}(1/\!\sqrt{2},0,i/\!\sqrt{2}), \\
\hspace{-1.2cm}m_{39} \hspace{-.3cm}&=&\hspace{-.3cm} - \frac{3}{2}(1+i)
[\tilde{b}_{33}(1,0,0) + \tilde{b}_{33}(0,0,1) ]
\nonumber \\
\hspace{-.3cm}&+&\hspace{-.3cm} 3\tilde{b}_{33}(1/\!\sqrt{2},0,1/\!\sqrt{2}) +
3i\tilde{b}_{33}(1/\!\sqrt{2},0,i/\!\sqrt{2}). \label{63of81}
\end{eqnarray}

Similarly, if Alice teleports the state 
$|\psi\rangle$ $=(|1\rangle+|2\rangle)/\sqrt{2}$  
($\alpha,\beta,\gamma=0,1/\sqrt{2},1/\sqrt{2}$) and the state 
$|\psi\rangle=$ $(|1\rangle+i|2\rangle)/\sqrt{2}$  
($\alpha,\beta,\gamma=0,1/\sqrt{2},i/\sqrt{2}$),  
Eqs.~(\ref{eqb11})-(\ref{eqb33}) give 
$m_{47},m_{48},m_{49},m_{57},m_{58},m_{59},m_{67},
m_{68},m_{69}$, the coeffici\-ents of the remaining 
block, namely, block $F_{3\times3}$,
\begin{eqnarray}
\hspace{-1.2cm}m_{47} \hspace{-.3cm}&=&\hspace{-.3cm} - \frac{3}{2}(1+i)
[\tilde{b}_{11}(0,1,0) + \tilde{b}_{11}(0,0,1) ]
\nonumber \\
\hspace{-.3cm}&+&\hspace{-.3cm} 3\tilde{b}_{11}(0,1/\!\sqrt{2},1/\!\sqrt{2}) +
3i\tilde{b}_{11}(0,1/\!\sqrt{2},i/\!\sqrt{2}), \label{64of81} \\
\hspace{-1.2cm}m_{48} \hspace{-.3cm}&=&\hspace{-.3cm} - \frac{3}{2}(1+i)
[\tilde{b}_{12}(0,1,0) + \tilde{b}_{12}(0,0,1) ]
\nonumber \\
\hspace{-.3cm}&+&\hspace{-.3cm} 3\tilde{b}_{12}(0,1/\!\sqrt{2},1/\!\sqrt{2}) +
3i\tilde{b}_{12}(0,1/\!\sqrt{2},i/\!\sqrt{2}), \\
\hspace{-1.2cm}m_{49} \hspace{-.3cm}&=&\hspace{-.3cm} - \frac{3}{2}(1+i)
[\tilde{b}_{13}(0,1,0) + \tilde{b}_{13}(0,0,1) ]
\nonumber \\
\hspace{-.3cm}&+&\hspace{-.3cm} 3\tilde{b}_{13}(0,1/\!\sqrt{2},1/\!\sqrt{2}) +
3i\tilde{b}_{13}(0,1/\!\sqrt{2},i/\!\sqrt{2}), \\
\hspace{-1.2cm}m_{57} \hspace{-.3cm}&=&\hspace{-.3cm} - \frac{3}{2}(1+i)
[\tilde{b}_{12}(0,1,0) + \tilde{b}_{12}(0,0,1) ]
\nonumber \\
\hspace{-.3cm}&+&\hspace{-.3cm} 3\tilde{b}_{12}(0,1/\!\sqrt{2},1/\!\sqrt{2}) +
3i\tilde{b}_{12}(0,1/\!\sqrt{2},i/\!\sqrt{2}), \\
\hspace{-1.2cm}m_{58} \hspace{-.3cm}&=&\hspace{-.3cm} - \frac{3}{2}(1+i)
[\tilde{b}_{22}(0,1,0) + \tilde{b}_{22}(0,0,1) ]
\nonumber \\
\hspace{-.3cm}&+&\hspace{-.3cm} 3\tilde{b}_{22}(0,1/\!\sqrt{2},1/\!\sqrt{2}) +
3i\tilde{b}_{22}(0,1/\!\sqrt{2},i/\!\sqrt{2}), \\
\hspace{-1.2cm}m_{59} \hspace{-.3cm}&=&\hspace{-.3cm} - \frac{3}{2}(1+i)
[\tilde{b}_{23}(0,1,0) + \tilde{b}_{23}(0,0,1) ]
\nonumber \\
\hspace{-.3cm}&+&\hspace{-.3cm} 3\tilde{b}_{23}(0,1/\!\sqrt{2},1/\!\sqrt{2}) +
3i\tilde{b}_{23}(0,1/\!\sqrt{2},i/\!\sqrt{2}), \\
\hspace{-1.2cm}m_{67} \hspace{-.3cm}&=&\hspace{-.3cm} - \frac{3}{2}(1+i)
[\tilde{b}_{31}(0,1,0) + \tilde{b}_{31}(0,0,1) ]
\nonumber \\
\hspace{-.3cm}&+&\hspace{-.3cm} 3\tilde{b}_{31}(0,1/\!\sqrt{2},1/\!\sqrt{2}) +
3i\tilde{b}_{31}(0,1/\!\sqrt{2},i/\!\sqrt{2}), 
\end{eqnarray}
\begin{eqnarray}
\hspace{-1.2cm}m_{68} \hspace{-.3cm}&=&\hspace{-.3cm} - \frac{3}{2}(1+i)
[\tilde{b}_{32}(0,1,0) + \tilde{b}_{32}(0,0,1) ]
\nonumber \\
\hspace{-.3cm}&+&\hspace{-.3cm} 3\tilde{b}_{32}(0,1/\!\sqrt{2},1/\!\sqrt{2}) +
3i\tilde{b}_{32}(0,1/\!\sqrt{2},i/\!\sqrt{2}), \\
\hspace{-1.2cm}m_{69} \hspace{-.3cm}&=&\hspace{-.3cm} - \frac{3}{2}(1+i)
[\tilde{b}_{33}(0,1,0) + \tilde{b}_{33}(0,0,1) ]
\nonumber \\
\hspace{-.3cm}&+&\hspace{-.3cm} 3\tilde{b}_{33}(0,1/\!\sqrt{2},1/\!\sqrt{2}) +
3i\tilde{b}_{33}(0,1/\!\sqrt{2},i/\!\sqrt{2}). \label{81of81}
\end{eqnarray}

Equations (\ref{1of81})-(\ref{81of81}) give all matrix elements of the sta\-te 
$\rho_{12}$ shared between Alice and Bob in terms of his single qutrit states at 
the end of nine different teleportation ``experiments''. Each one of the nine states 
with Bob are associated with one of the nine states listed above that Alice might 
choose to implement the teleportation protocol, namely, 
\begin{eqnarray}
&|0\rangle,|1\rangle,|2\rangle,
(|0\rangle +|1\rangle)/\sqrt{2},(|0\rangle +i|1\rangle)/\sqrt{2}, \nonumber \\
&(|0\rangle +|2\rangle)/\sqrt{2}, (|0\rangle +i|2\rangle)/\sqrt{2}, \nonumber \\
&(|1\rangle +|2\rangle)/\sqrt{2},(|1\rangle +i|2\rangle)/\sqrt{2}. 
\label{9states}
\end{eqnarray}

Analogously to the two-qubit case \cite{rig26}, the nine states
that Alice should teleport to Bob for a complete reconstruction of $\rho_{12}$
are not unique. Instead of $|\psi\rangle=(|0\rangle+|1\rangle)/\sqrt{2}$  and 
$|\psi\rangle=(|0\rangle+i|1\rangle)/\sqrt{2}$, for example, Alice can use 
$|\psi\rangle=(|0\rangle-|1\rangle)/\sqrt{2}$  and 
$|\psi\rangle=(|0\rangle-i|1\rangle)/\sqrt{2}$. However, Alice must employ at least 
nine states for the teleportation-based QST protocol to work. As already noted 
above, each one of these nine states is related to a specific density matrix that describes Bob's qutrit after the teleportation protocol finishes. These density 
matrices have nine independent real parameters (the three real diagonal coefficients 
and the complex off-diagonal ones). Therefore, nine different teleported states will provide $9\times 9 = 81$ real parameters. This is exactly the number of independent parameters that characterizes an arbitrary two-qutrit quantum state. 
Furthermore, Eqs.~(\ref{1of81})-(\ref{81of81}) are
valid for low rank matrices too \cite{rig26}. 
For low rank matrices, many 
coefficients of $\rho_{12}$ are zero. The experimental signature that $\rho_{12}$
is a low rank matrix (lower than $9$ for two qutrits) is associated with 
$b_{jj}(\alpha,\beta,\gamma)$ (or equivalently $\tilde{b}_{jj}(\alpha,\beta,\gamma)$)
being zero for certain values of $j,\alpha,\beta$ and $\gamma$ 
[see Eqs.~(\ref{1of81})-(\ref{27of81})]. 
In other words, for certain states teleported by Alice,
some of the matrix coefficients related to Bob's state at the end of the teleportation protocol are zero. For very low rank matrices, we will also have $Q_j(\alpha,\beta,\gamma)$ equal to zero for particular values of $j,\alpha,\beta$ and $\gamma$  [see Eq.~(\ref{q1})].

\section{Reconstructing one-qutrit}

\begin{figure}[!ht]
\centering
\includegraphics[width=4cm]{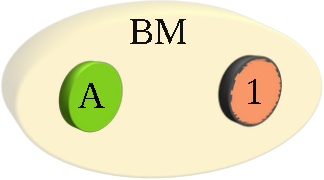}
\caption{\label{fig2}(color online) 
The one-qutrit Bell-state-measurement-only reconstruction protocol. The picture shows
schematically the key ingredient to its functioning: Bell measurements. By properly
projecting the unknown qutrit $\varrho_1$ (orange disk) and a set of known qutrits 
$\rho_A$ (green disk) onto Bell states, we can express the matrix elements of 
$\varrho_1$ as a function of the probabilities to measure Bell states. 
See text for more details.  
}
\end{figure}

The previous two-qutrit protocol works if Bob can fully reconstruct single qutrits.
At the end of each one of the nine teleportation ``experiments'', Bob 
has to determine the density matrix describing his qutrit [see Eq.~(\ref{stepE})].
This can be done using any standard QST protocol \cite{vog89,leo95,whi99,jam01,ari01,bri04,moh06,moh07,moh08,cra10,gro10,tot10,chr12,bau13,bau13b,lan17,tor18} or by using the Bell-state-measurement-only single qutrit 
reconstruction protocol explained below, where only Bell measurements (BMs) are the required tools to reconstruct a qutrit. 
This means that only BMs and the ability to generate the nine states (\ref{9states})
are all that is needed to fully determine one and two-qutrit states 
(see Fig. \ref{fig2}).

Let us now present the Bell-state-measurement-only single qutrit 
reconstruction protocol. The arbitrary qu\-trit we want to determine is written as 
\begin{equation}
\varrho_{1} =  
\left(
\begin{array}{ccc}
a_{11} & a_{12} & a_{13} \\
a_{12}^* & a_{22} & a_{23} \\
a_{13}^* & a_{23}^* & a_{33}
\end{array}
\right).
\label{stepF}
\end{equation}
The equivalent now to Eq.~(\ref{stepA}) is $\rho=\rho_A\otimes \varrho_1$. Making 
a BM on the two-qutrit state $\rho$, we realize that the probability to get the Bell
state $|B_j\rangle$ is
\begin{eqnarray}
\hspace{-.5cm}Q_1(\alpha,\beta,\gamma) &=&\frac{1}{3}[ 
a_{11}|\alpha|^2 + a_{22}|\beta|^2 + a_{33}|\gamma|^2 \nonumber \\
&+& 2\,\Re (a_{12}\alpha \beta^* + a_{13}\alpha \gamma^* + a_{23}\beta \gamma^*)], \label{q1b}\\
\hspace{-.5cm}Q_2(\alpha, \beta, \gamma) &=& Q_1 \left( \alpha, e^{i \frac{4\pi}{3}} \beta, e^{i \frac{2\pi}{3}} \gamma \right), \\
\hspace{-.5cm}Q_3(\alpha, \beta, \gamma) &=& Q_1 \left( \alpha, e^{i \frac{2\pi}{3}} \beta, e^{i \frac{4\pi}{3}} \gamma \right),\\
\hspace{-.5cm}Q_4(\alpha, \beta, \gamma) &=& Q_1 \left( \gamma, \alpha, \beta \right),\\
\hspace{-.5cm}Q_5(\alpha, \beta, \gamma) &=& Q_1 \left( \gamma, e^{i \frac{4\pi}{3}} \alpha, e^{i \frac{2\pi}{3}} \beta \right),\\
\hspace{-.5cm}Q_6(\alpha, \beta, \gamma) &=& Q_1 \left( \gamma, e^{i \frac{2\pi}{3}} \alpha, e^{i \frac{4\pi}{3}} \beta \right),\\
\hspace{-.5cm}Q_7(\alpha, \beta, \gamma) &=& Q_1 \left( \beta, \gamma, \alpha \right),\\
\hspace{-.5cm}Q_8(\alpha, \beta, \gamma) &=& Q_1 \left( \beta, e^{i \frac{4\pi}{3}} \gamma, e^{i \frac{2\pi}{3}} \alpha \right),\\
\hspace{-.5cm}Q_9(\alpha, \beta, \gamma) &=& Q_1 \left( \beta, e^{i \frac{2\pi}{3}} \gamma, e^{i \frac{4\pi}{3}} \alpha \right). \label{q9b}
\end{eqnarray}
Note that the relations among the several probabilities above are the same 
we observed for the two-qutrit case [Eqs.~(\ref{q1})-(\ref{q9})]. Therefore, similarly
to the two-qutrit protocol, we now fix our attention to the case in which Alice measures the Bell state $|B_1\rangle$. 

Observing Eq.~(\ref{q1b}), we see that if $\rho_A$ is the density matrix representing
$|0\rangle, |1\rangle$, and $|2\rangle$, we get $a_{11},a_{22}$, and $a_{33}$ as functions of the probability of measuring the Bell state
$|B_1\rangle$. In the same fashion, if $\rho_A$ is given by 
$(|0\rangle + |1\rangle)/\sqrt{2}$ and $(|0\rangle + i|1\rangle)/\sqrt{2}$, we can solve
the two linear equations to get $a_{12}$. Finally, if $\rho_A$ is 
$(|0\rangle + |2\rangle)/\sqrt{2}$ and $(|0\rangle + i|2\rangle)/\sqrt{2}$, we obtain $a_{13}$ and if $\rho_A$ is 
$(|1\rangle + |2\rangle)/\sqrt{2}$ and $(|1\rangle + i|2\rangle)/\sqrt{2}$, we fix
$a_{23}$. Putting it differently, all coefficients of the matrix $\varrho_1$ are  functions of the probabilities to measure the Bell state $|B_1\rangle$ when Alice 
projects each one of the nine states available to her and the state to be determined 
onto the Bell basis. 

Being more explicit,
\begin{eqnarray}
\hspace{-.5cm}a_{11} \hspace{-.2cm}&=&\hspace{-.2cm} 3 Q_{1}(1,0,0),\\
\hspace{-.5cm}a_{22} \hspace{-.2cm}&=&\hspace{-.2cm} 3 Q_{1}(0,1,0), \\
\hspace{-.5cm}a_{33} \hspace{-.2cm}&=&\hspace{-.2cm} 3 Q_{1}(0,0,1), \\ 
\hspace{-.5cm}a_{12} \hspace{-.2cm}&=&\hspace{-.2cm} -\frac{3}{2}(1+i)[Q_{1}(1,0,0)+Q_{1}(0,1,0)] \\ 
\hspace{-.2cm}&+&\hspace{-.2cm}3Q_{1}(1/\sqrt{2},1/\sqrt{2},0)+3iQ_{1}(1/\sqrt{2},i/\sqrt{2},0), \nonumber \\
\hspace{-.5cm}a_{13} \hspace{-.2cm}&=&\hspace{-.2cm} -\frac{3}{2}(1+i)[Q_{1}(1,0,0)+Q_{1}(0,0,1)]\\ 
\hspace{-.2cm}&+&\hspace{-.2cm}3Q_{1}(1/\sqrt{2},0,1/\sqrt{2})+3iQ_{1}(1/\sqrt{2},0,i/\sqrt{2}), \nonumber \\
\hspace{-.5cm}a_{23} \hspace{-.2cm}&=&\hspace{-.2cm} -\frac{3}{2}(1+i)[Q_{1}(0,1,0)+Q_{1}(0,0,1)] \\ 
\hspace{-.2cm}&+&\hspace{-.2cm}3Q_{1}(0,1/\sqrt{2},1/\sqrt{2})+3iQ_{1}(0,1/\sqrt{2},i/\sqrt{2}). \nonumber
\label{single}
\end{eqnarray}
It is worth mentioning that Alice must be able to prepare the same set of nine states
[Eq.~(\ref{9states})] used in the two-qutrit teleportation-based QST protocol for the one-qutrit reconstruction protocol to work. 
This feature will repeat itself when we deal with three or more qutrits. Also,
there is no need to prepare states where the three vectors spanning the standard basis
are simultaneously present. In other words, the more difficult to prepare superpositions such as 
$(|0\rangle+|1\rangle+|2\rangle)/\sqrt{3}$ are not needed.

\section{Reconstructing three or more qutrits}

\begin{figure}[!ht]
\centering
\includegraphics[height=2.75cm,width=5.25cm]{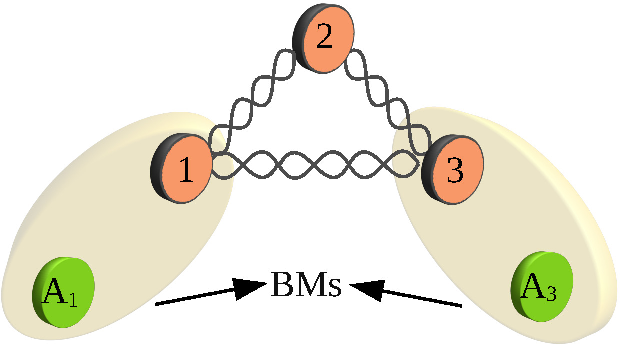}
\caption{\label{fig3}(color online) 
The three-qutrit teleportation-based QST protocol. Two input states $\rho_{A_1}$ and
$\rho_{A_3}$ (green disks) 
are prepared by Alice. Each input state, together with 
one qutrit of the state we want to determine (orange disks), are projected 
onto Bell states as shown in the figure. Bob then is informed by Alice of the results of those BMs and of the input states she prepared (qutrits $A_1$ and $A_3$).
With these pieces of information Bob is sure that his qutrit is  
$\varrho_2$ [Eq.~(\ref{stepE3})], from which he can reconstruct $\rho_{123}$,
the unknown three-qutrit state (orange disks). See text for details.
}
\end{figure}

Before presenting the general scheme leading to the $n$-qutrit teleportation-based QST protocol, it is worthwhile analyzing in greater detail the three-qutrit case
(see Fig. \ref{fig3}). 
This will give us a more solid understanding of how things change as we scale up for 
higher values of $n$.

For instance, in the three-qutrit scenario, 
Eq.~(\ref{stepA}) changes to 
\begin{eqnarray*}
\hspace{-.5cm}\rho \hspace{-.2cm}&=&\hspace{-.2cm} \rho_{A_{1}} \otimes \rho_{123} \otimes \rho_{A_{3}} \nonumber \\
\hspace{-.5cm}&=&\hspace{-.2cm} 
\left(\hspace{-.1cm}
\begin{array}{ccc}
|\alpha|^2 & \alpha\beta^* & \alpha\gamma^* \\
\alpha^*\beta & |\beta|^2 & \beta\gamma^* \\
\alpha^*\gamma & \beta^*\gamma & |\gamma|^2
\end{array}
\hspace{-.1cm}\right)\hspace{-.05cm} 
\otimes \rho_{123}  \otimes \hspace{-.05cm}
\left(\hspace{-.1cm}
\begin{array}{ccc}
|\delta|^2 & \delta\varepsilon^* & \delta\zeta^* \\
\delta^*\varepsilon & |\varepsilon|^2 & \varepsilon\zeta^* \\
\delta^*\zeta & \varepsilon^*\zeta & |\zeta|^2
\end{array}
\hspace{-.1cm}\right).
\end{eqnarray*}
Alice now has to teleport a pair of input states given by 
$|\psi\rangle_{A_1}=\alpha|0\rangle + \beta|1\rangle+ \gamma|2\rangle$
and $|\psi\rangle_{A_3}=\delta|0\rangle + \varepsilon|1\rangle + \zeta|2\rangle$, 
where $|\alpha|^2+|\beta|^2+|\gamma|^2=1$ and $|\delta|^2+|\varepsilon|^2+|\zeta|^2=1$.
The three-qutrit state we want to determine is given by $\rho_{123}$.

Alice then implements two independent BMs. One on qutrits $1$ and $A_1$ and another 
on qutrits $3$ and $A_3$, informing Bob of her measurement outcomes (which Bell states 
she obtained). Bob's qutrit, in its turn, is given by the most natural extension of Eq.~(\ref{stepE}),
where two indexes instead of one are needed to correctly label the $81$ possible
outcomes of the two BMs at Alice's,
\begin{equation}
\varrho_{2} =  \frac{Tr_{A_1,1,A_3,3}[P_{i,j} \rho P_{i,j}]}{Q_{i,j}}
=\left(
\begin{array}{ccc}
b_{11} & b_{12} & b_{13} \\
b_{12}^* & b_{22} & b_{23} \\
b_{13}^* & b_{23}^* & b_{33}
\end{array}
\right). 
\label{stepE3}
\end{equation}
In Eq.~(\ref{stepE3}) we have that $Q_{i,j}$ denotes Alice's chance to
measure the Bell states $|B_i\rangle$ and $|B_j\rangle$ 
while $P_{i,j}$ is the projector
describing those two independent BMs:
\begin{eqnarray}
Q_{i,j} &=& Tr[{P_{i,j} \rho}], \label{prob3} \\
P_{i,j} &=& P_{i} \otimes \mathbb{1} \otimes P_{j}, 
\end{eqnarray}
with $P_i(P_j)$ defined at Eq.~(\ref{projectorB}).

At the end of the teleportation protocol, Bob's single qutrit state will be a 
function of $\rho_{123}$ and of the two states teleported by Alice. Using 
$9\times 9=81$ pairs of inputs obtained by combining the states shown in
Eq.~(\ref{9states}), we get the exact number of equations whose solutions 
give all the matrix elements of $\rho_{123}$ as functions of $b_{ij}(\alpha,\beta,\gamma,\delta,\varepsilon,\zeta)$, 
i.e., as functions of Bob's states at the end of the $81$ 
different teleportation protocols that can be executed with $81$ different pairs
of inputs.

Let us dig a little deeper of why the three-qutrit protocol works as shown in 
Fig.~\ref{fig3}. For definiteness, we fix our attention to the case in which 
Alice measures the Bell states $|B_1\rangle_{A_1,1}|B_1\rangle_{A_3,3}$.  In the 
three-qutrit protocol, the $81$ possible pairs of states that Alice should teleport to Bob give a total of $81$ different density matrices with Bob at the end of those 
$81$ ``experiments'' where Alice measures $|B_1\rangle_{A_1,1}|B_1\rangle_{A_3,3}$.  Noting that each density matrix with Bob has nine independent real parameters, we get 
$9\times 81 = 729$ independent real parameters with Bob. But these $729$ real parameters are associated with $729$ linear equations that when solved give 
the $729$ real parameters that characterizes $\rho_{123}$. Note that an arbitrary
three-qutrit density matrix is a $27\times 27$ matrix. And since it is Hermitian,
it has $27^2=729$ independent real parameters, exactly the same 
number of independent parameters with Bob.

The extension to the the $n$-qutrit case is somewhat direct. Now, Alice 
teleports $n-1$ single qutrits, employing all $n-1$ arrangements of states
that can be formed from the basic nine ones listed in Eq.~(\ref{9states}).
Each qutrit with Alice is paired with a qutrit of the $n$-qutrit state to be
reconstructed and then $n-1$ independent BMs are made. The remaining qutrit of 
the $n$-qutrit state is the one Bob should determine after the teleportation finishes.
The information contained
in the several density matrices describing the remaining qutrit, after each one of
the distinct teleportations of $n-1$ qutrits
from Alice to Bob, is what allows him to infer the original state describing the 
multipartite $n$-qutrit state.

Being more specific, we first note that the density matrix describing an $n$-qutrit 
state has $3^n\times 3^n$ elements. Due to Hermiticity, the $n$ diagonal elements 
are real with only half of the complex off-diagonal elements independent. 
We thus have $9^n$ independent real parameters characterizing the $n$-qutrit state. 
Now, since we have nine different 
basic single qutrit states to choose in order to compose the $(n-1)$-qutrit state
to be teleported from Alice to Bob, we have a total of $9^{n-1}$ different
$(n-1)$-qutrit states to be teleported. Fixing our attention, for definiteness, in a 
particular sequence of BM results attained by Alice, each teleportation leads to a
specific density matrix with Bob at the end of the teleportation protocol. 
But an arbitrary single qutrit state has nine real parameters and thus Bob will have $9\times 9^{n-1} = 9^n$ real independent parameters, all functions of the $n$-qutrit state he wants to determine. This number of independent parameters is precisely the 
number of independent real parameters fully describing an $n$-qutrit density matrix. 
To finally obtain the matrix elements of the unknown $n$-qutrit state in terms of 
measured quantities (his single qutrits matrix elements), 
Bob must solve a linear system of $9^n$ equations.

\section{Conclusion}

We extended to qutrits (spin-1 systems) the teleporta\-tion-based quantum state 
tomography (QST) protocol originally developed for qubits 
(spin-1/2 systems) \cite{rig26}. Similarly to the qubit protocol, the basic ingredient
to its implementation is the spin-1 extension of Bell measurements (BMs). 

We showed that if Alice is able to prepare nine basic pure single qutrit states
[see Eq.~(\ref{9states})],
nine teleportation ``experiments'', where those known states are teleported from Alice to Bob, suffice to determine the unknown state of an arbitrary two-qutrit state.
At the end of each teleportation experiment, Bob will end up with a single qutrit state. The knowledge of those nine single qutrit states at the end of each teleportation 
experiment allows him to reconstruct the two-qutrit state shared with Alice.

We also explained how to extend the present protocol to reconstruct an $n$-qutrit 
multipartite state and also how to use BMs alone to determine the density matrix
of a single qutrit. The latter protocol, dubbed Bell-state-measurement-only single qutrit reconstruction protocol, allows us to determine the density matrix of a 
single qutrit by just measuring the probabilities to obtain a given Bell state 
after nine different BM experiments. Each BM 
is implemented by projecting onto the spin-1 Bell basis each one of the nine basic states alluded to above and the qutrit whose density matrix we want to know.

Similar to the qubit teleportation-based QST protocol \cite{rig26}, the qutrit teleportation-based QST protocol can be implemented ``remotely'', with the $n$-partite qutrit state spread into $n$ different locations. In all locations but one the parties must perform a BM and then inform the remaining party of their measurement
results and which states they teleported. Working together in this way, the remaining party can fully reconstruct the $n$-qutrit state as previously described in this work.

We also showed that the number of different teleportation experiments one must do to implement the present
protocol scales exponentially with $n$, 
similar to the scaling of the resources needed to implement the standard 
QST protocols \cite{vog89,leo95,whi99,jam01,ari01,bri04,moh06,moh07,moh08,cra10,gro10,tot10,chr12,bau13,bau13b,lan17,tor18}. However, this scaling is based on the assumption that 
we have no prior information about the $n$-qutrit state we want to reconstruct.
It remains an open question to determine 
the minimum number of teleportations needed when one 
deals with particular but important type of states, such as matrix product states \cite{fan92,gar07} or almost pure states, which are known to require less resources to
be reconstructed via standard QST techniques \cite{cra10,gro10,bau13,bau13b,lan17}.
Another important topic to investigate is the connection of the present protocol to quantum process tomography \cite{chu97} using the mathematical framework of invertible and non-invertible maps given in Ref. \cite{jag22}.
And it also remains open whether or not the ideas contained here and in 
Ref. \cite{rig26} can be extended to continuous-variable systems
\cite{vai94,bra98,fur98,har07,rig08}.

\section*{Acknowledgements}
GR thanks the Brazilian agency CNPq (National Council for
Scientific and Technological Development) for funding. 



\end{document}